\documentclass[9pt, conference]{IEEEtran}
\IEEEoverridecommandlockouts

\usepackage{cite}
\usepackage{amsmath,amssymb,amsfonts,bbding}
\usepackage{algorithm}
\usepackage{algorithmic}
\usepackage{graphicx}
\usepackage{textcomp}
\usepackage{xcolor}
\usepackage{siunitx}
\usepackage{tipa}
\usepackage{hyperref}
\newcommand{\ourmethod}{VADUSA}
\def\BibTeX{{\rm B\kern-.05em{\sc i\kern-.025em b}\kern-.08em
    T\kern-.1667em\lower.7ex\hbox{E}\kern-.125emX}}
\usepackage{multirow}
\usepackage{subcaption}  

\begin{document}
 
\title{Fast and High-Quality Auto-Regressive Speech Synthesis \\ via Speculative Decoding}

\author{
\IEEEauthorblockN{Bohan Li, Hankun Wang, Situo Zhang, Yiwei Guo, Kai Yu$^{\dagger}$\thanks{$^{\dagger}$Kai Yu is the corresponding author.}}
\IEEEauthorblockA{
    \textit{MoE Key Lab of Artificial Intelligence, AI Institute;}\\
    \textit{
     X-LANCE Lab, Department of Computer Science and Engineering,} \\
     \textit{Shanghai Jiao Tong University,
     }Shanghai, China \\
\{everlastingnight, wanghankun, situozhang, yiwei.guo, kai.yu\}@sjtu.edu.cn}

}

\maketitle

\begin{abstract}

The auto-regressive (AR) architecture, exemplified by models such as GPT, is extensively utilized in modern Text-to-Speech (TTS) systems. However, it often leads to considerable inference delays, primarily due to the challenges associated with next-token prediction in long speech sequences. In this work, we introduce VADUSA, one of the first approaches to accelerate AR-based TTS through speculative decoding. Our findings demonstrate that VADUSA not only delivers a significant reduction in inference time but also enhances TTS quality by employing draft heads to predict future speech tokens in an auto-regressive manner. Additionally, the incorporation of a tolerance mechanism during the sampling process further boosts performance, yielding approximately a 3 $\times$ speedup in AR TTS. Moreover, our approach exhibits strong generalization across diverse datasets and various speech token types.
\end{abstract}


\vspace{-4pt}
\section{Introduction}
  
Large language models (LLMs) with auto-regressive (AR) architectures~\cite{NEURIPS2020_1457c0d6-gpt3, touvron2023llama} have gained significant success in recent years. They generate text byte-pair encoding (BPE) tokens using a next-token prediction strategy \cite{Vaswani2017AttentionIA}. This strategy samples the next token from a multinomial distribution generated based on the history tokens, which is simple but effective in building coherence over a long context. With the invention of discrete neural audio tokens~\cite{hsu20221hubert, baevski2020wav2vec, vq-wav2vec, du2022vqtts, defossez2022high-EnCodec}, a speech utterance can also be coded as a discrete token sequence. The paradigm in LLMs is then introduced to the speech modeling and synthesis field. Studies like SPEAR-TTS~\cite{kharitonov2023speartts}, VALL-E~\cite{Wang2023NeuralCL} and BASE-TTS~\cite{lajszczak2024base} and other works~\cite{du2024vall, Song2024ELLAVSN} implement decoder-only AR architectures for speech codec generation with large-scale training data, allowing for generating natural-sounding speech. 

However, there is a significant difference between text and speech data: the sequence of speech tokens is often much longer than that of text tokens for the same sentence. For example, a piece of 10-second speech needs 500 HuBERT tokens (\SI{50}{\hertz})~\cite{hsu20221hubert} to represent, while the text transcription of the same utterance only costs approximately 20 to 40 text BPE tokens. It is reasonable because the text is more information-dense, while speech captures finer acoustic details over time, resulting in a longer token sequence. Moreover, the trivial AR architecture predicts only one token at each inference step, causing speech generation time especially long. This creates a challenge between the need for low-latency spoken language synthesis and modeling long speech sequences. 

To mitigate the issue, one way of previous efforts is to reduce the input/output sequence length via distillation or better information compression, i.e., build low bitrate speech tokens. A bunch of works on neural codec~\cite{ticodec, singlecodec, ji2024wavtokenizer} attempt to lower the bitrate via a nicely configured vector quantize (VQ) module and a well-designed training process. Another inspiring research is aBPE~\cite{Shen2023AcousticBF, li2024effectiveness}, which applies BPE to discrete speech tokens, achieving lossless compression and showing big potential in accelerating speech synthesis. However, these low-bitrate tokens either compromise speech reconstruction accuracy or make the frameshift variant, affecting the naturalness and quality of speech synthesis. 

Another way is to predict more speech tokens at one decoding step, which is the main focus of this paper. This idea has developed since the RNN era, started by Subscale WavRNN~\cite{Kalchbrenner2018EfficientNA}. It folds the wav token sequence in a subscaling way to realize predicting multiple tokens in parallel. Recently, VALL-E 2~\cite{Chen2024VALLE2N} uses a chunk-wise manner to generate a chunk of future tokens (typically 2 tokens) in one AR iteration. However, such methods usually lead to quality degradation because the history condition is insufficient when predicting at least half of a sequence. In contrast, speculative decoding \cite{Leviathan2022FastIF}, which is widely used in LLM inference, provides acceleration without performance loss by simply incorporating an additional draft model. 


In light of the potential of speculative decoding, we apply the advanced speculative decoding strategy, MEDUSA \cite{Cai2024MedusaSL}, to an AR TTS model, VALL-E \cite{Wang2023NeuralCL}, conducting comprehensive experiments on various discrete tokens. In attempting to do it, we observed significant challenges. The complexity and variability inherent in speech synthesis often resulted in suboptimal prediction accuracy and acceptance rates, making it difficult to achieve the desired acceleration and quality. This prompted us to innovate beyond mere adaptation, leading to the development of the {\ourmethod} method. By integrating a tolerance mechanism, {\ourmethod} not only accelerates the decoding process but also enhances the overall synthesis robustness and quality by effectively managing the speculative decoding's inherent uncertainty. This novel approach proves crucial in bridging the gap between fast decoding and maintaining high synthesis fidelity.

\vspace{-1pt}
\section{{\ourmethod}: Fast and High-Quality AR TTS}

In this section, we elaborate {\ourmethod}, a novel AR TTS decoding method that simultaneously achieves fast decoding and high-quality speech synthesis. The section first introduces MEDUSA~\cite{Cai2024MedusaSL}, an effective and lossless speculative decoding method for AR transformer-based models.
Then we dig into an unsatisfactory vanilla combination of VALL-E and MEDUSA and analyze where its limitations come from.
After that, incorporating the tolerance mechanism, we propose {\ourmethod}, a non-trivial integration for MEDUSA and VALL-E, realizing both decoding acceleration and synthesis quality improvements. Finally, we briefly describe the design of TTS-oriented sparse candidate trees. 
\begin{figure}[htbp]
    \centering
    \label{Fig:Framework}
    \begin{subfigure}[b]{0.2\textwidth}
        \centering
        \includegraphics[width=\textwidth]{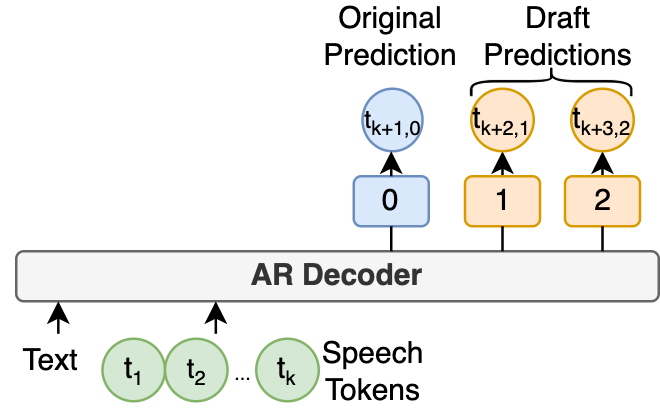}
        \caption{Prefill the model with transcript text and prompt speech tokens.}
        \label{fig:sub_a}
    \end{subfigure}
    \hfill
    \begin{subfigure}[b]{0.25\textwidth}
        \centering
        \includegraphics[width=\textwidth]{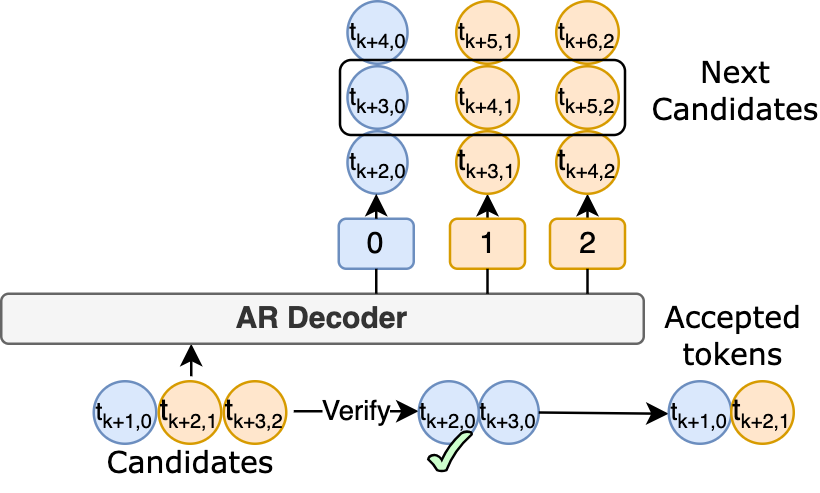}
        \caption{Speculative decoding and vanilla verification process.}
        \label{fig:sub_b}
    \end{subfigure}
    \hfill
    \begin{subfigure}[b]{0.25\textwidth}
        \centering
        \includegraphics[width=\textwidth]{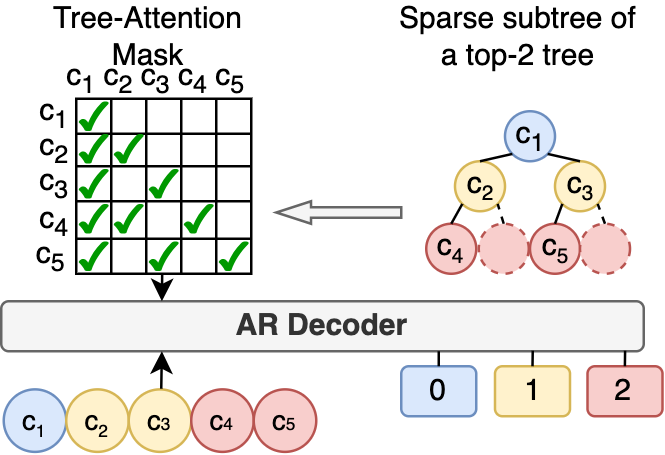}
        \caption{Structures of Sparse Tree and Tree-Attention mask.}
        \label{fig:sub_c}
    \end{subfigure}
    \hfill
    \caption{The overview framework of vanilla VADUSA.}
\end{figure}

\subsection{Speculative Decoding and MEDUSA}
Speculative decoding~\cite{Leviathan2022FastIF} is an efficient, lossless technique for accelerating AR decoding. The key idea is to use a small, fast draft model to generate predictions, which are then verified by the target model in parallel. To avoid the expensive process of training a separate draft model, MEDUSA~\cite{Cai2024MedusaSL} introduces several additional draft heads on top of the original model. We directly implement this method on the vanilla \ourmethod.
Specifically, as demonstrated in Fig.\ref{fig:sub_a}, in the prefilling pass with the prompt text and speech sequence $\{t_1,t_2,...,t_k\}$, say the original prediction head predicts the first next token $t_{k+1,0}$, then the $i$-th draft head is responsible for predicting the $(i+1)$-th next token $t_{k+(i+1),i}(i \ge 1)$ by the supervision of a cross-entropy loss multiplied with $\lambda^i, \lambda\in(0,1)$, enabling the generation of multiple tokens in parallel. The generated first next token and draft tokens will be concatenated in order and input into the same model for the next forward pass. In the decoding pass, the model will verify the draft tokens almost in parallel. For example, $t_{k+2,1}$ will be \textit{accepted} iff. it matches the `correct token' sampled from the distribution produced by the original head of $t_{k+2,0}$. Similarly, $t_{k+3,2}$ will be accepted iff. $t_{k+2,1}$ is accepted and it matches the original prediction $t_{k+3,0}$. As shown in Fig.~\ref{fig:sub_b}, if $t_{k+2,1}$ is accepted but $t_{k+3,2}$ is rejected, the original head continues from $t_{k+3,0}$, which is the prediction of the last accepted token, appending new draft predictions $t_{k+4,1}$ and $t_{k+5,2}$ as candidates for the subsequent decoding pass. In this scenario, the \emph{acceptance length} for the current pass is 2.

To make better `guesses', MEDUSA also considers multiple candidate tokens for each head and constructs a draft tree where the root node represents the first next token, each node a candidate token, and each root path a candidate continuation. For each draft head, the tokens with top-$k$ logits are chosen as candidates. Each non-leaf node has $k$ children. The $(i+1)$-th layer has $k^{i-1}$ nodes and is filled with the top-$k$ tokens produced by the $i$-th draft head. All the root-to-leaf paths form all possible combinations of the candidates. With a carefully constructed tree-attention mask, the whole tree of tokens can be verified within one forward pass (named tree attention).
This approach significantly improves both the acceptance length of drafts and the extra computational cost is also reasonable by facilitating \textit{sparse trees} that will be introduced in Section \ref{subsec:sparse-tree}. The simplicity of MEDUSA draft heads and the compatibility with any transformer-based AR model make MEDUSA highly adaptable and efficient.

{
}

\subsection{Vanilla Combination and Re-thinking}
In the vanilla version of {\ourmethod}, we simply use a VALL-E TTS model as a base model for speech synthesis and integrate several MEDUSA heads on top of it to speed up generation and implement a kv-cache for in-context memory. However, in initial experiments, we found its acceptance length is not comparable to the results of MEDUSA on text LLMs. This is because speech tokens such as those generated by systems like EnCodec~\cite{defossez2022high-EnCodec}, HuBERT~\cite{hsu20221hubert}, and wav2vec 2.0~\cite{baevski2020wav2vec}, are obtained through vector quantization or clustering of IDs, rather than having distinct abstract semantic differences like text tokens~\cite{Choi2024SelfSupervisedSR}. Under the AR language modeling paradigm, multiple speech tokens often have very similar statistical features. This means that the multinomial distributions predicted by models like VALL-E tend to be more `average' and selecting the highest probability top-$k$ tokens using draft heads only covers a small probability range. Given that during validation the original head samples only once as the `correct token', it is challenging for the draft heads in VALL-E to make accurate predictions.

Moreover, from the perspective of improving sample quality, tree attention can be seen as an efficient version of beam search. In beam search, by foreseeing tokens of the next few steps, we can select the combination with the highest probability to improve the sample quality. Similarly, under tree attention, we use draft heads to foresee the future tokens, and we hope the model accepts a candidate path with a higher accept probability.


\begin{algorithm}[t]
\caption{Verification with tolerance mechanism}
\label{alg:1}
\begin{algorithmic}[1]
\REQUIRE AR decoder $D$, number of draft heads $k$, a branch of candidates $\mathcal{B}:=\{c_0, c_1, ..., c_k\}$, tolerance $\tau$.
\STATE $\mathcal{L} \gets D(\mathcal{B})$
\FOR{$i = 0,1,...,k$}
\STATE $v_i \gets multinomial\_sampling(top\_p(\mathcal{L}(i)), \tau)$
\ENDFOR
\STATE $a_0 = c_0$
\STATE $k^* = k$
\FOR{$i = 0,1,...,k-1$}
\IF{$c_{i+1} \in v_{i}$}
\STATE $a_{i+1} = c_{i+1}$
\ELSE
\STATE $k^* = i$
\STATE \textbf{break}
\ENDIF
\ENDFOR
\RETURN $\mathcal{A} \gets (a_0, a_1, ..., a_{k^*})$
\end{algorithmic}
\end{algorithm}

\subsection{{\ourmethod} Decoding with the Tolerance Mechanism}
We propose {\ourmethod} decoding method with \textit{tolerance mechanism} to help the VALL-E model achieve a higher acceleration rate and better synthesis quality concurrently. The tolerance mechanism allows the original head to sample multiple times, which means the `correct tokens' is also multiple. The number of original head's samplings is controlled by the tolerance value $\tau$.
Alg. \ref{alg:1} demonstrates the verification process of one branch of candidates on the top-$k$ tree with tolerance mechanism.
Between the accepted results on the top-$k$ branches, we first select the longest sequences. Among those, we choose the one with the highest acceptance probability. 
In other words, if nodes on the left side of the top-$k$ tree correspond to smaller $k$ values during its construction, we prioritize results from the leftmost branches when the lengths are equal. In this way, on the one hand, with a slightly higher cost of the verification process, the acceptance length is considerably increased and finally leads to a higher real acceleration rate. On the other hand, by choosing the best candidate path among all accepted paths, the sample quality is improved compared to a trivial single-time sampling strategy.

Note that {\ourmethod} can still be compatible with any improved sampling strategy. For instance, methods such as avoiding sequential repetition in VALL-E 2 \cite{Chen2024VALLE2N} and constraining the attention window in ACI \cite{Cai2024MedusaSL} can be integrated simply by taking the sampling outcomes of these strategies as the predictions of the original head. If multiple tolerances are present, we can sample multiple times using the aforementioned strategies. This ensures that while the quality of sampling is maintained, acceleration effects are also achieved. In our experiments, we will evaluate using only the normal sampling method.

\subsection{TTS-Oriented Sparse Tree Design}
\label{subsec:sparse-tree}
As demonstrated in Fig.\ref{fig:sub_c}, we simply visualize the sparse tree and its tree-attention mask of a top-$2$ tree In practice, $k$ is usually set as 10, i.e., each non-leaf candidate node has 10 children, which leads to an exponential explosion in the node number of the full candidate tree. To reduce the computational cost of tree attention, we need to design a \textit{sparse tree} that only contains a tiny part of the candidate tree. Within a constrained number of nodes, we expect the sparse tree to help the model make as many acceptances as possible. Therefore, the design of sparse trees depends on the choice of calibration dataset and the output distribution of draft heads (otherwise, we cannot obtain the acceptance probability). Unfortunately, the existing trees provided by MEDUSA are based on text datasets and BPE tokenizers, which may not fit in speech synthesis scenarios. 

Using the criterion of maximizing the expectation of accepted length, we use the off-the-shelf greedy algorithm in MEDUSA~\cite{Cai2024MedusaSL} to build a TTS-oriented sparse tree. Specifically, for each type of discrete audio token we use, we separately train a VADUSA model. Then, we run forward passes with the full tree on a subset of LibriTTS~\cite{zen2019libritts} to obtain the accepted probability of each node. Finally, we run the greedy algorithm to build the sparse tree for each token type.

\setlength{\tabcolsep}{5pt} 
\begin{table}[!t]
\caption{Performance on AR TTS inference. ``Tuned'' refers to whether fine-tune the base model({\ourmethod}-wot or {\ourmethod}-wt) and ``Acc." refers to the mean number of accepted tokens.}
\begin{center}\scalebox{0.96}{
\begin{tabular}{cccccc}
\hline
\textbf{Settings}                                                                                  & \textbf{Tuned} & \textbf{WER$\downarrow$}(\%) & \textbf{UTMOS$\uparrow$} & \textbf{Speedup$\uparrow$} & \textbf{Acc. $\uparrow$} \\ \hline
\multirow{2}{*}{w/o. {\ourmethod}}                                                         & \XSolidBrush        & 7.61& 4.34 $\pm$ 0.014   & -       & -         \\
                                                                                          & \CheckmarkBold           & 7.29    & 4.35 $\pm$ 0.013    & -       & -         \\ \hline
\multirow{2}{*}{4$\times$ {\ourmethod}}                                                          & \XSolidBrush          & 7.21    & 4.35 $\pm$ 0.014    & 2.37$\times$& 2.85      \\
                                                                                          & \CheckmarkBold          & 6.56    & 4.35 $\pm$ 0.014    & 2.43$\times$& 3.00      \\ \hline
\multirow{2}{*}{\begin{tabular}[c]{@{}c@{}}4$\times$ {\ourmethod} \\ + $\tau$=3\end{tabular}} & \XSolidBrush          & 5.94    & \textbf{4.36 $\pm$ 0.013}     & 2.89$\times$& 3.80      \\
                                                                                          & \CheckmarkBold          & \textbf{5.71}    & \textbf{4.36 $\pm$ 0.013}    & \textbf{2.94$\times$}& \textbf{3.87}      \\ \hline
\end{tabular}}
\end{center}
\label{table1}
\end{table}

\section{Experiments}
All experiments are conducted on NVIDIA A800-SXM4-80GB GPUs, including training with different strategies and evaluation on both objective and subjective metrics. Our implementation is adapted from the open-source repositories\footnote{https://github.com/lifeiteng/vall-e}\footnote{https://github.com/FasterDecoding/MEDUSA}.
\subsection{Experimental Setups}
\subsubsection{Architecture of TTS system and draft heads}
The whole system is cascaded by tokenizers of text and speech, a codec language model, and a speech-codec-based vocoder. The codec language model, which is the main part of the system, consisted of 12 transformer layers, with 16 attention heads with independent kv-caches, 1024 hidden dimensions and 4096 feed-forward dimensions. We utilize a grapheme-to-phoneme converter as the text tokenizer.
We tokenize speech with 2048 k-means clustering on features from the last layer of a HuBERT-large \cite{hsu20221hubert} model. Two additional speech tokens are used in the ablation study: wav2vec 2.0 \cite{baevski2020wav2vec} with its 2-group vector quantization using IDs of top 16000 statistically occuring frequency groups on LibriTTS \cite{zen2019libritts}, and EnCodec \cite{defossez2022high-EnCodec} with its 8-layer residual vector quantization of 1024 size per codebook. All models used in tokenizers are pretrained. The vocoder, CTX-vec2wav \cite{Du2023UniCATSAU}, is constructed by two conformer blocks with 2 layers and 184 attention dimensions and a HifiGAN \cite{Kong2020HiFiGANGA}, trained for converting semantic tokens to waveform. {\ourmethod} draft heads are connected to the last transformer layer of the codec language model, constructed by a residual block with a linear layer and a SiLU activation layer \cite{Elfwing2017SigmoidWeightedLU}.

\subsubsection{Training Settings}
We perform experiments on both LibriHeavy \cite{kang2024libriheavy} with 50k hours of English speech data and LibriTTS \cite{zen2019libritts} with 585 hours. For the codec language model, we train two versions as the base model on them separately as foundational training stage: one epoch on Libriheavy or 20 epochs on LibriTTS,  where the learning rates are 0.05 and the warm-up steps are set as 200. Then we trained {\ourmethod} heads on LibriTTS for 10 epochs with a fixed base model, mentioned as {\ourmethod}-wot(without tuning) and fine-tuning base model, mentioned as {\ourmethod}-wt(with tuning), with $\lambda$ set as 0.8, 40 of warm-up steps and 0.002 of learning rates. For the vocoder, replicas of CTX-vec2wav \cite{Du2023UniCATSAU} are trained for senmatic tokens of HuBERT and wav2vec2.0 on LibriTTS for 1 million steps.

\subsubsection{Configurations in decoding strategy}

We apply the nuclear sampling method with top-$p$, setting the temperature to 0.9 and 1.0, on both the original head of the base model and the draft heads of {\ourmethod}. The number of draft heads is set to 4 or 6. For the draft tree, the top-$k$ value is fixed at 10 to define the branching factor, and 64 nodes are selected as candidates by default, with 96 and 128 nodes used for ablation studies. Candidate expectations were calculated using over 3000 audio samples, each longer than 6 seconds, from the LibriTTS dataset to construct the sparse tree. For the tolerance mechanism, we set $\tau$ values ranging from 1 to 4 across all configurations.

\subsection{Evaluations}

The exploration of inference speedup is conducted by measuring the number of discrete tokens generated per second during AR inference in the codec language model. All metrics are measured on UniCATS Testset B\cite{Du2023UniCATSAU}. The speedup ratio is calculated by dividing the output rate of the model equipped with {\ourmethod} heads by that of the baseline model. Additionally, the number of the mean accepted tokens is used to assess the effectiveness of the {\ourmethod} decoding with the tolerance mechanism. 
TTS performance is evaluated using word error rates (WER) measured by a conformer-transducer model\footnote{{https://huggingface.co/nvidia/stt\_en\_fastConformer\_transducer\_large}}. 
We also use the UTokyo-SaruLab MOS (UTMOS) prediction system\footnote{https://github.com/sarulab-speech/UTMOS22} to evaluate synthesis quality of the generated speech objectively. As shown in Table \ref{table1}, the experimental settings varies in three key aspects: (1) whether 4 {\ourmethod} heads or only base model participates during inference; (2) {\ourmethod}-wot or {\ourmethod}-wt is used; and (3) whether the tolerance mechanism is applied by setting $\tau$=3.

The results suggest that {\ourmethod} performs well in AR TTS systems. Inference acceleration is substantial, with no degradation in generation quality—in fact, performance improved when the tolerance mechanism is applied or when the base model is tuned. This improvement is particularly noticeable in models trained on smaller datasets, where the {\ourmethod} heads help capture in-context information. This behavior contrasts with the application of similar strategies in LLMs for natural language modeling tasks.

\subsection{Ablation Study}

In this section, we present results from various configurations of the proposed methods. Furthermore, we demonstrate their overall effectiveness on large public datasets and different types of discrete speech tokens, which can serve as a valuable reference for future implementations. 

\subsubsection{Effectiveness analysis of {\ourmethod} configurations and tolerance mechanism}

\begin{figure}[!t]
\centerline{\includegraphics[width=0.44\textwidth]{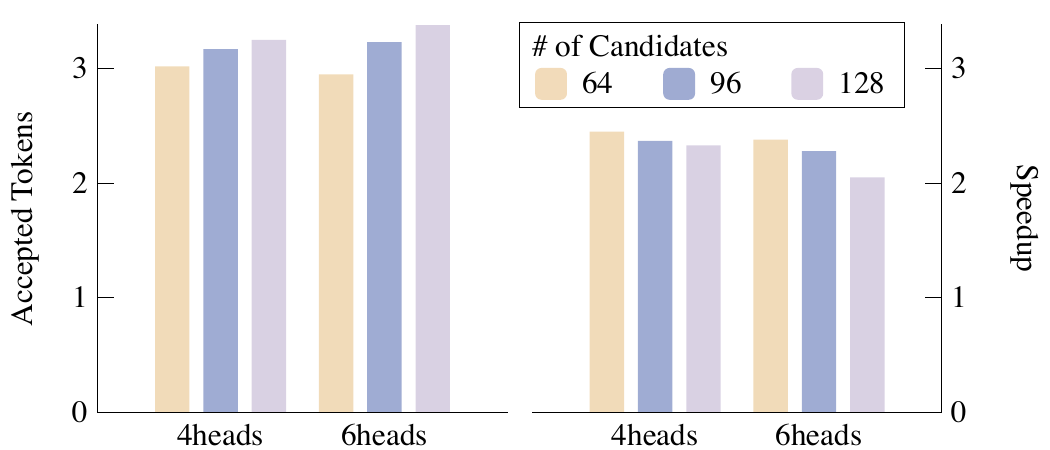}}
\caption{Performances of different VADUSA configurations, with the same base model trained on HuBERT+k-means2048 tokens.}
\label{fig1}
\end{figure}

\begin{figure}[!t]
    \centering
    \includegraphics[width=0.48\textwidth,trim=0 0 0 0,clip]{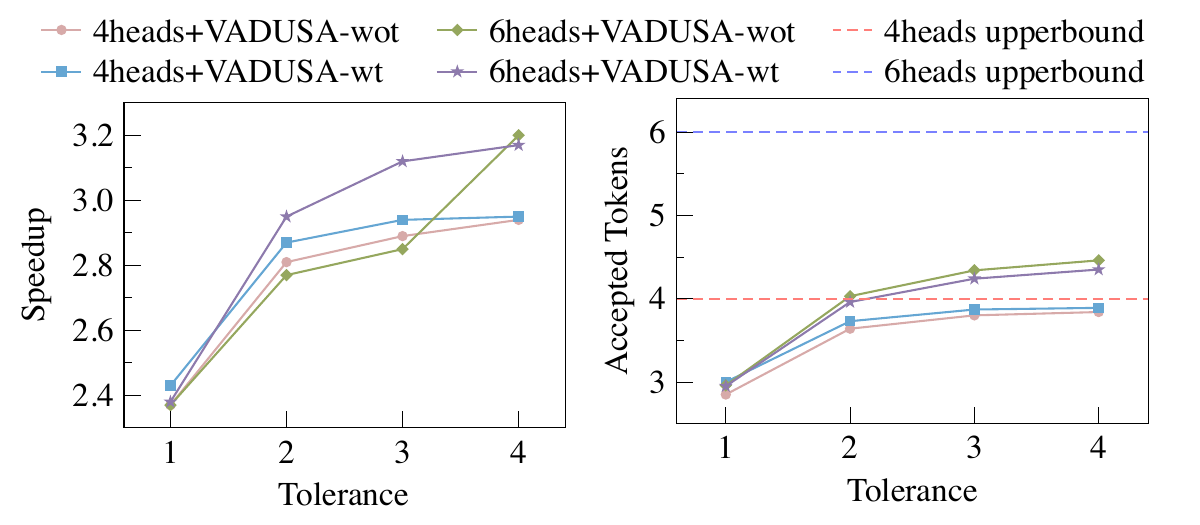}
    \caption{Speedup performance and mean number of accepted tokens in increasing number set in tolerance strategy, with base model trained on HuBERT+k-means 2048 tokens, selecting 64 candidates per decoding step.}
    \label{fig2}
\end{figure}

We adjust the candidate length to represent the number of selected nodes in sparse tree construction. The results, shown in Fig. \ref{fig1}, reveal performance variations of {\ourmethod}-wt with draft heads set to {4, 6} and candidate choices set to {64, 96, 128}. Contrary to expectations, increasing the number of candidates does not result in improved speedup performance, due to the increased computational cost associated with accepting a higher number of candidates. The tolerance mechanism is varied between 1 and 4, with 4 or 6 heads trained using either the {\ourmethod}-wot or {\ourmethod}-wt strategy. As illustrated in Fig. \ref{fig2}, the tolerance mechanism performs well in general {\ourmethod} configurations, with a larger tolerance value leading to better acceleration.

\subsubsection{Performance of different speech tokens}

To demonstrate the effectiveness of the methods across different tokens, we conduct experiments on both semantic and acoustic tokens. For semantic tokens, we utilize HuBERT-extracted tokens with 2048 k-means clusters, as well as wav2vec2.0-extracted tokens with the top-16000 frequency in LibriTTS of a combined 2-group codebook. For acoustic tokens, we employ the EnCodec model, pre-trained with 50Hz RVQ and 8 codebooks of size 1024. The results presented in Tab. \ref{table2} indicate that {\ourmethod} training strategies and the proposed tolerance mechanism are adaptable to both large-codebook semantic tokens and acoustic tokens. However, due to the large codebook size of wav2vec2.0 and the remaining seven codebooks of EnCodec, the logit distributions from the prediction heads are relatively flat, resulting in low prediction accuracy and consequently reduced acceptance rates for draft heads. This explains the suboptimal performance of decoding speedup observed in both cases. Adjusting the top-$k$ parameter during sparse tree construction may offer a potential solution.

\begin{table}[!t]

\caption{Results of base-model-only inference and inference with best-performing strategies({\ourmethod}-wt strategy with 4 draft heads and $\tau$=3) in different token types.}
\label{table2}
\begin{center}
\begin{tabular}{ccccc}
\hline
\textbf{Tokens}                              & \textbf{Strategies}                                                                                                        & \textbf{WER$\downarrow$(\%)} & \textbf{Speedup$\uparrow$}       & \textbf{Acc.$\uparrow$}   \\ \hline
\multicolumn{1}{c|}{HuBERT 2048}    & \multicolumn{1}{c|}{\multirow{3}{*}{\begin{tabular}[c]{@{}c@{}}{\ourmethod}-wt\\ base-model-only\end{tabular}}}                 & 7.29                & \textbf{-}     & -             \\
\multicolumn{1}{c|}{wav2vec2.0 16k} & \multicolumn{1}{c|}{}                                                                                           & 5.71                & -              & -             \\
\multicolumn{1}{c|}{EnCodec 1024}   & \multicolumn{1}{c|}{}                                                                                           & 3.91                & -              & -             \\ \hline
\multicolumn{1}{c|}{HuBERT 2048}    & \multicolumn{1}{c|}{\multirow{3}{*}{\begin{tabular}[c]{@{}c@{}}{\ourmethod}-wt\\ + 4$\times$heads, \\ + 3$\times$ tolerance\end{tabular}}} & 5.71                & \textbf{2.94$\times$} & \textbf{3.87} \\
\multicolumn{1}{c|}{wav2vec2.0 16k} & \multicolumn{1}{c|}{}                                                                                           & 3.48                & 1.30$\times$         & 1.89          \\
\multicolumn{1}{c|}{EnCodec 1024}   & \multicolumn{1}{c|}{}                                                                                           & \textbf{3.17}       & 1.85$\times$         & 1.96          \\ \hline
\end{tabular}
\end{center}
\end{table}

\subsubsection{Generality in large dataset}

Our strategies also demonstrate generalizability on a scaled-up dataset, LibriHeavy, which contains approximately 50,000 hours of audiobook audio. Aside from the Word Error Rates (WERs), the performance of the proposed methods is consistent with that observed on the smaller LibriTTS dataset. The discrepancy in WERs is understandable, as the capacity of additional heads to learn in-context information has inherent limitations. This reflects a trade-off that arises when relying on draft heads within the constraints of the base model.

\begin{table}[!t]
\label{table3}
\caption{The effectiveness of the proposed methods in scaled-up dataset. We compare the performance on settings of 4-head fine-tuned or not {\ourmethod} model({\ourmethod}-wt or {\ourmethod}-wot) with $\tau$=3.}
\begin{center}
\begin{tabular}{ccccccc}
\hline
\textbf{Dataset}                     & \textbf{Tuned}              & \textbf{WER$\downarrow$(\%)}  & \textbf{UTMOS$\uparrow$}         & \textbf{Speedup$\uparrow$}       & \textbf{Acc.$\uparrow$}      \\ \hline
\multirow{2}{*}{LibriTTS}   & \XSolidBrush     & 5.94  & 4.36 $\pm$ 0.013       & 2.89$\times$          & 3.80 \\ 
                            & \CheckmarkBold       & 5.71  & 4.36 $\pm$ 0.013       & 2.94$\times$ & 3.87 \\ \hline
\multirow{2}{*}{LibriHeavy} & \XSolidBrush     & \textbf{2.65} & 4.35 $\pm$ 0.013        & 2.94$\times$          & \textbf{3.88} \\ 
                            & \CheckmarkBold       & 2.85  & \textbf{4.36 $\pm$ 0.012}       & \textbf{2.95$\times$} & \textbf{3.88} \\ \hline
\end{tabular}
\end{center}
\end{table}
 
\section{Conclusion}

In conclusion, VADUSA, inspired by the integration of VALL-E and MEDUSA, demonstrates impressive effectiveness in accelerating AR TTS decoding and enhancing speech quality. This is due to the draft heads' ability to learn in-context information, expanding the capabilities of TTS models. This approach benefits AR models like VALL-E, which often exhibit instability in speech synthesis. By selecting the most appropriate tokens in just a few decoding steps, VADUSA reduces the likelihood of collapse caused by incorrect token selection. The proposed tolerance mechanism further enhances this effect. Moreover, experimental results highlight the method's generalizability, offering valuable insights for future exploration and implementation.

\section*{Acknowledgment}

This work was supported by the China NSFC Project (No. 92370206), the Key Lab of Suzhou on Linguistic Computing (SZS2024005) and Shanghai Municipal Science and Technology Major Project (2021SHZDZX0102).


\newpage
\bibliographystyle{IEEEtran}
\bibliography{refs}

\end{document}